\documentstyle[prl,twocolumn,aps]{revtex}

\input{psfig}
\def\apj #1 #2 #3 {#1, ApJ, {\bf #2}, #3}
\def\apjl #1 #2 #3 {#1, ApJ, {\bf #2}, L#3}
\def\apjs #1 #2 #3 {#1, ApJS, {\bf #2}, #3}
\def\aap  #1 #2 #3 {#1, A\&A, {\bf #2}, #3}
\def\mnras #1 #2 #3 {#1, MNRAS, {\bf #2}, #3}
\def\pra #1 #2 #3 {#1, Phys.~Rev.~A., {\bf #2}, #3}
\def\prb #1 #2 #3 {#1, Phys.~Rev.~B., {\bf #2}, #3}
\def\prc #1 #2 #3 {#1, Phys.~Rev.~C., {\bf #2}, #3}
\def\prd #1 #2 #3 {#1, Phys.~Rev.~D., {\bf #2}, #3}
\def\pre #1 #2 #3 {#1, Phys.~Rev.~E., {\bf #2}, #3}
\def\prl #1 #2 #3 {#1, Phys.~Rev.~Lett., {\bf #2}, #3}
\def\plb #1 #2 #3 {#1, Phys.~Lett.~B., {\bf #2}, #3}
\def\science #1 #2 #3 {#1, Science., {\bf #2}, #3}
\def\nature #1 #2 #3 {#1, Nature., {\bf #2}, #3}
\def\nphysa #1 #2 #3 {#1, Nucl.~Phys.~A., {\bf #2}, #3}
\def\nphysb #1 #2 #3 {#1, Nucl.~Phys.~B., {\bf #2}, #3}
\def\nphysbs #1 #2 #3 {#1, Nucl.~Phys.~B.~Suppl., {\bf #2}, #3}

\def\h#1{\hbox{${}^{#1}$H}}

\def\h502{\hbox{$ h^{2}_{50}$}}

\def\fun#1#2{\lower3.6pt\vbox{\baselineskip0pt\lineskip.9pt
  \ialign{$\mathsurround=0pt#1\hfil##\hfil$\crcr#2\crcr\sim\crcr}}}
\begin{document}
\bigskip
\bigskip
%

\title{Big Bang Nucleosynthesis with a New Neutron Lifetime}
\author{
G. J. Mathews$^{1}$,
T. Kajino$^{2,3}$,
T. Shima$^4$
}
\address{$^1$University of Noter Dame, Center for Astrophysics, Notre Dame,
IN 46556
}
\address{$^2$National Astronomical Observatory, Mitaka, Tokyo 181-8588, Japan
}
\address{
$^3$Department of Astronomy, Graduate School of Science,
University of Tokyo, 7-3-1
Hongo, Bunkyo-ku, Tokyo 113-0033, Japan }
\address{$^4$Research Center for Nuclear Physics,
 Osaka University,
 Osaka 567-0047, Japan
}
\maketitle

\date{\today}
\begin{abstract}
We show that
the predicted primordial helium production is significantly reduced
when new measurements of the neutron lifetime and the implied
enhancement in the weak reaction rates are included in big-bang
nucleosynthesis.
Therefore, even if a narrow
uncertainty in the observed helium abundance is adopted,
this brings the
constraint on the baryon-to-photon ratio from
BBN and the observed helium into better accord with
the independent determination of the baryon content
deduced from the {\it WMAP} spectrum of power fluctuations in the
cosmic microwave background,
 and measurements of primordial deuterium in
narrow-line quasar absorption systems at high redshift.
\end{abstract}
\section{Introduction}
Big-bang nucleosynthesis
(BBN) plays a crucial role in constraining cosmological
models.  It is essentially the only probe of physics in the early universe
during the interval from $\sim 1-10^4$ sec in the radiation dominated epoch.
As such, it is important to have accurate predictions of the light
element abundances produced in this era.

 The single unknown parameter for standard BBN is the baryon-to-photon ratio
during the nucleosynthesis epoch.  All light abundances are
a simple function of this parameter.
In this regard, it has been noted for some time
\cite{Wagoner,Wagoner2,YTSSO,Malaney,SKM,Burles,Cyburt}
that the nucleosynthesis yields from the big bang are particularly sensitive
to the neutron lifetime which
affects BBN in two ways.
For one, changing the neutron lifetime $\tau_n$  implies  different weak
reaction rates through the relation between the neutron lifetime and
the weak coupling constant.
\begin{equation}
\tau_n^{-1} = \frac{G_F^2}{2 \pi^3} (1 + 3 g_A^2) m_e^5 \lambda_0~~,
\end{equation}
where $G_F$ is the Fermi coupling constant and $g_A$ is the
axial-vector coupling of the nucleon.  The quantity $m_e$ is the electron mass
and $\lambda_0$ is the phase-space integral for neutron decay.
To a good approximation, weak reactions cease once the
weak reaction rate
\begin{equation}
\Gamma = (7 /60)\pi (1 + 3 g_A^2)G_F^2 T^5~~,
\end{equation}
becomes smaller than the Hubble expansion rate.
\begin{equation}
H \approx [(8/3) \pi G \rho_\gamma]^{1/2}~~,
\end{equation}
 where $\rho_\gamma
= (\pi^2 /30) g_* T^4$ is the energy density in relativistic particles,
and $g_*$ the total number of effectively massless degrees of freedom
at the relevant epoch.

 Equating these two rates gives
the freezeout temperature, $T_f \approx 1$ MeV.
Once weak reactions freeze out
the ratio of the number of neutrons/protons remains fixed at the freezeout
value except for neutron decay.
However, changing the neutron half life
resets the temperature $T_f$ at which
weak-reactions freeze out.

For example, a shorter lifetime for neutron decay means that
the reaction rates remain greater than the Hubble expansion rate
until a lower freezeout temperature. This shifts the
equilibrium neutron-to-proton ratio at freezeout.
To a good approximation  this $n/p$ ratio
is just given by thermal equilibrium to be
$n/p = \exp{\{-\Delta m/T_f\}}$,
where $\Delta m$ is the mass difference between the neutron and the proton.
Since most of the neutrons
remaining until the nucleosynthesis epoch at $t \sim 200$ sec
are converted to $^4$He, there is a simple approximate
 relation between
the $n/p$ ratio at freezeout and the helium mass fraction from BBN
\begin{equation}
Y_p \approx  2n/(n + p) = 2(n/p)/(n/p+1) ~~,
\end{equation}
where $n$ and $p$ refer to the number densities of neutrons
and protons, respectively.
The other dependence of $Y_p$ on
the neutron lifetime simply comes from the fact that
some neutrons can decay in the interval between weak freezeout
($t \sim 1$ sec) and nucleosynthesis ($t \sim 200$ sec).
Taken together, both of these effects imply that the shorter
the neutron lifetime, the lower the predicted BBN helium
abundance.  

Regarding the neutron lifetime, it is of particular interest that a new and
very accurate measurement of the neutron lifetime has recently been reported 
\cite{Serebrov} using ultracold neutrons in a gravitational trap. 
There are several distinguishing features of this measurement.
Among the most important  are: 1) that it involves
the best-measured storage time ($872 \pm 1.5$ sec) of neutrons in
the trap; 
 2) the ability to measure the spectrum of the ultracold neutrons after they have been stored in the trap; and 3) an improvement in the coating of the traps which improves the reliability for the different geometries.  
These features are particularly important improvements because all measurements of the neutron lifetime  involve an ambiguity between the  neutron decay lifetime $\tau_n$
 and the storage lifetime $\tau_{storage}$ in the trap.
To circumvent this, it is necessary to conduct a number of
measurements with different storage lifetimes which can be then extrapolated
to zero storage loss rate ($\tau_{storage}^{-1} \rightarrow 0$)
 to determine the decay rate due to neutron decay alone.
This is a source of considerable systematic error.     The present result
is a major improvement in previous extrapolations in that the neutron
storage loss rate was not only accurately measured but was as much as a factor of two smaller than the best previous
measurements, making the inferred neutron decay lifetime
much less subject to systematic error.  Indeed, the difference between
 the best-measured storage time and the inferred neutron lifetime is
 only 5 s, whereas in the previous best measurements the extrapolation 
was made over an interval of 105 s and therefore less reliable.

The neutron lifetime deduced by this method
is significantly reduced to
$878.5 \pm 0.7_{stat} \pm 0.3_{sys}$ sec.  This value differs from the previous  mean weighted 
world average of $885.7 \pm 0.8$ \cite{pdg}  by six standard deviations.
It differs  
from the previous most precise  result of $885.4 \pm 0.9_{stat} \pm 0.3_{sys}$ \cite{Arzumanov}
by four standard deviations.   Indeed, including the new result
of Ref. \cite{Serebrov} into deriving a new weighted mean world average according to
the methods of the {\it Particle Data Group} \cite{pdg} reduces the
mean weighted world average by over four standard deviations
to $881.9 \pm 0.6$.   More conservatively, however, this
 weighted mean uncertainty may not be appropriate due to the inconsistency among the data.  A chi-squared minimization instead of a weighted mean gives a larger uncertainty of $\pm 1.6$ s,
 which we adopt here. This larger uncertainty and smaller lifetime together help to bring the primordial helium abundance into concordance
 with other determinations of the baryon-to-photon ratio
 as described below.

On the other hand, there are reasons to consider the new value by itself, independently of the previous measurements.  In addition to
the vast improvement  in the present measurement,
another aspect which lends particular credibility to this new result is
the fact that when this new lifetime is used as a unitarity
test of the Cabibbo-Kobayashi-Maskawa (CKM) matrix \cite{Serebrov},
together with the current value of the $\beta$-asymmetry in neutron decay
\cite{betaasymm}, there is excellent agreement with
the standard-model predictions.
Such is not the case for the current world average \cite{Serebrov}.

 If this new
lower value for the neutron lifetime is adopted as a most extreme case,
 then this substantially reduces the expected $^4$He abundance from
  primordial nucleosynthesis.
This is particularly important for BBN cosmology as we now explain.

\section{Light Element Abundances}
One of the powers of BBN is that all of the light element
abundances are determined in terms of a single parameter 
$\eta_{10}$
which is the baryon-to-photon ratio in units of $10^{-10}$.
The crucial test of the standard BBN is, therefore,  whether a single
value of $\eta_{10}$ can be found which reproduces all of the
observed primordial abundances.
The different light element abundances are determined by different
means. This makes each determination an important independent
check on BBN.

Primordial deuterium is best determined from its absorption line
in  high redshift Lyman $\alpha$ clouds.
The average of measurements of six absorption-line systems
towards five QSOs gives
 \cite{Kirkman}
$D/H = 2.78^{+0.44}_{-0.38} \times 10^{-5}$.
This would imply an value of  $\eta_{10} = 5.9 \pm 0.5$.
This is an important result because it is also very close to the value
$\Omega_b h^2 = 0.0224 \pm 0.0009$ ($\eta_{10} = 6.13 \pm 0.25$)
deduced \cite{WMAP} from the {\it WMAP} independent determination of the
baryon content at the epoch of photon last scattering.
Because of the concordance of these two independent methods,
the {\it WMAP} determination of $\eta_{10}$ is generally
accepted as the most  accurate determination.

The primordial lithium abundance , on the other hand, is
inferred from old low-metallicity halo stars.
Such stars exhibit an approximately constant (``Spite plateau'')
lithium abundance as a function of surface temperature.
This is taken to be the primordial abundance.
There is, however, some controversy \cite{pinns,vauclair} concerning
the depletion of $^7$Li on the surface of such halo stars
and/or during the big bang itself \cite{Coc}.
 For the
present purposes we adopt the
value from \cite{Ryan00}
$^7$Li$=
1.23^{+0.68}_{-0.32}\times 10^{-10}$, where the errors are
95\% confidence limits.

The primordial helium abundance is obtained by measuring
extragalactic $\rm H{II}$ regions in low-metallicity irregular
galaxies.  Often in the past,
the primordial helium abundance $Y_p$ so deduced
 tended to reside in one of two possible values (a low value, e.g.
$Y_p = 0.238 \pm 0.002 \pm 0.005$,  \cite{Fields} and a high value $Y_p  0.2452
\pm 0.0015$ \cite{Izatov}).
There is also, however,  a current dilemma
regarding the uncertainty in the observationally  determined
primordial helium abundance.
 Many recent
evaluations (e.g. \cite{Izatov}) give a rather narrow range
of abundance uncertainty.  For our purposes we adopt the value
of \cite{Izatov} as a representative result.
 On the other hand, the extent of systematic
errors in these analyses is still being debated.  Another recent
study \cite{Olive} has adopted a more conservative approach
and concluded that correlations in various
uncertainties could stretch the error in the inferred primordial abundance.  Their represantitve analyses 
yields  $Y_p = 0.249 \pm 0.009 $ and they argue in favor of
range of allowed values of $0.232 \le Y_p \le 0.258$.

While this is being sorted out, however,  it
has been deduced by several authors (cf. \cite{Coc,Cyb03})
that the combined deuterium and {\it WMAP} constraints on
the baryon-to-photon ratio
implies that the primordial helium  abundance should be
$Y_p = 0.2484^{0.0004}_{0.0005} $ \cite{Cyb03} or $Y_p = 0.2479 \pm 0.0004$
 \cite{Coc}.

If we adopt the narrow helium abundance of \cite{Izatov}
and the {\it WMAP} constraint of \cite{Cyb03} there
is, therefore,  a possible  2-3$\sigma$ discrepancy between the
$^4$He + $^7$Li and the D + {\it WMAP} results.
This dilemma with regards to BBN is depicted by
dashed lines on  Figure \ref{fig:1}.

  One of course could (and probably should) disregard this dilemma if the uncertainty is
as large as deduced in \cite{Olive}. However, if this dilemma is real,
then it may provide insight into new physics beyond the minimal BBN model,
for example, brane-world effects \cite{Ichiki03},
cosmic quintessence \cite{Yahiro}, time varying constants \cite{Ichikawa}, etc.
\cite{Malaney}.
In this paper we point out an important result, however, that even if the
most narrow uncertainty in the deduced primordial helium is adopted,
then a significant portion of the discrepancy between BBN and
the CMB results can be accounted for simply by adopting the new neutron
lifetime.

\section{Results}
For illustration of the implications of the new value for the neutron lifetime, we have made calculations of standard homogenous big bang nucleosynthesis for three values of the neutron lifetime.  These are: 1)  the previous world average ($885.7 \pm 0.8$ sec);
2) the new world average ($881.9 \pm 1.6$ sec) which includes the new measurement of Ref.~\cite{Serebrov}; 
and 3) the newest lower value
of \cite{Serebrov} ($878.5 \pm 0.7 \pm 0.3$ sec).  
 
 The benchmark code used for the present illustration 
 is the standard big bang nucleosynthesis code
originally developed by Wagoner \cite{Wagoner} and made user friendly by Kawano \cite{Kawano}.  This code is available for public download \cite{code}.  The reaction rates and uncertainties are those adopted in \cite{SKM}.  Although newer reaction rate compilations and uncertainties have been evaluated
\cite {Burles,Cyburt,Descouvemont}, this code is readily available and adequate for the benchmark comparison of interest here.

\begin{figure}[htb]
\psfig{figure=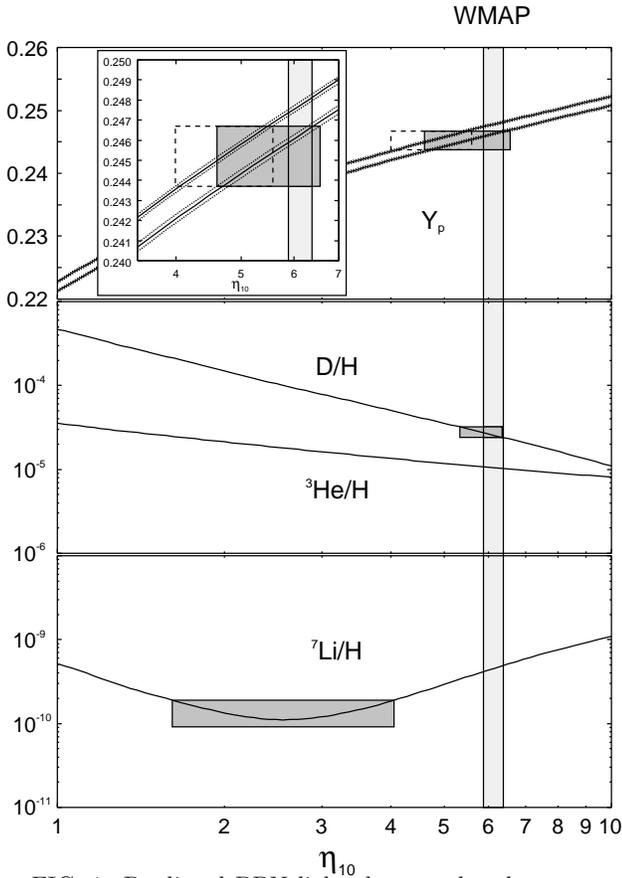,width=3.2in}
\caption{  Predicted BBN light-element abundances 
vs. the baryon-to-photon ratio $\eta_{10}$ in units
of $10^{-10}$.  
These are compared
with the observationally inferred \protect\cite{Coc} primordial abundances (horizontal lines) and the independent
determination of $\eta_{10}$ from the {\it WMAP} results 
(light shaded region).
The top box shows the primordial helium abundances.  
The insert shows an expanded view of 
$Y_p$ near the allowed region.  
The banded regions indicate the range of predicted  
$Y_p$ due to the neutron lifetime uncertainty.  
The upper  lines are  based upon the previous world average 
$\tau_n = 885.7 \pm 0.8$ s.
The lower lines are based upon the new measured value of
$\tau_n = 878.5 \pm 0.8$ s.
The previous allowed $\eta_{10}$  
values (shown by the dashed open box) shifts to
the dark shaded box  if the new neutron lifetime is adopted.
 }
\label{fig:1}
\end{figure}

  Figure \ref{fig:1} compares the primordial nucleosynthesis yields based upon both the previously adopted world average with the
yields based upon  the new neutron lifetime measurement and its uncertainty.  The insert shows an expanded view of the primordial
helium abundance for $\eta_{10}$ values near those allowed by the
various observational constraints.

From this figure it is clear that the primary effect of altering the neutron lifetime is to lower the primordial helium abundance prediction.  The uncertainty in the predicted $Y_p$ is indicated by parallel bands on the figure. The uncertainty 
remains the nearly same with the new lifetime because the uncertainty 
in the new neutron lifetime is nearly the same as that of the previous world average.  Though not shown on this figure, the uncertainty in predicted $Y_p$ increases by a factor of $\approx 1.5$  if the larger error ($\pm 1.6$ s) in the new world average is adopted.

 The effect on other light elements is so small ($^<_\sim 1\%$) as to be  indiscernible from the line widths on the figure.  
The key point of Figure \ref{fig:1} is that
now the primordial helium abundance required for the
baryon-to-photon ratio given in the {\it WMAP} and/or D/H QSO absorption-line
results reduces from
$Y_p = 0.2479 \pm 0.0006$ to $Y_p = 0.2463 \pm 0.0006$ when
using the new value for new neutron lifetime.   For 
comparison, incorporating the new lifetime measurement into a new mean world average would require
$Y_p = 0.2470 \pm 0.0009$.  These two later values overlap (within $1\sigma$) with the  uncertainty of even the narrower of the observationally inferred helium abundance 
 \cite{Izatov}  of $Y_p = 0.2452 \pm 0.0015$.  
 
 Alternatively,
 the $\eta_{10}$ values implied by an observed helium abundance
 of $Y_p = 0.2452 \pm 0.0015$, are $5.5 \pm 0.9$ for the new lifetime,
 or $5.1 \pm 1.1$ for the new weighted mean lifetime as compared
to $4.8 \pm 0.8$ based upon the previous world average.  These
are to be compared with the {\it WMAP} + D/H determination of 
$\eta_{10} = 6.13 \pm 0.25$.
Hence, even with this small correction to the neutron lifetime, and  adopting a narrow range for the observational uncertainty in  $Y_p$,
the implied $\eta_{10}$ value for either the new lifetime or new weighted average now overlaps
the value required by the {\it WMAP} and D/H analysis.
This significantly further constrains nonstandard models for BBN,
and strengthens the viability of standard BBN as a probe of cosmology.

Of course, one must still deal with the problem of $^7$Li
overproduction in BBN which will have to be resolved
by $^7$Li destruction, either within the big bang itself \cite{Coc}
or during subsequent stellar evolution \cite{pinns,vauclair}.
We also emphasize that there is still additional uncertainty in the
BBN production of
helium and other light-element abundances due to uncertainties in
nuclear reaction rates, particularly the d(p,$\gamma$)$^3$He, d(d,n)$^3$He,
d(d,p)$^3$H, and $^3$He(a,$\gamma$)$^7$Be.

\begin{acknowledgments}
We acknowledge the valuable help of Satoshi Kawanomoto in the preparation
of this manuscript.  One of the authors (TS) would like to acknowledge useful
conversations with Prof. M. Utsuro (RCNP, Osaka Univ.) regarding details
of the various neutron lifetime measurements.
Work at the University of Notre Dame is supported
by the U.S. Department of Energy under
Nuclear Theory Grant DE-FG02-95-ER40934.
Work at NAOJ supported in part by the Grants-in-Aid for
Scientific Research (12047233, 13640313, 14540271) and for
Specially Promoted Research (13002001)
of the Ministry of Education, Science, Sports, and Culture of Japan.
\end{acknowledgments}

\end{document}